\documentclass[12pt]{article}
\usepackage{amssymb}
\usepackage{amsmath}
\usepackage[unicode,bookmarks,bookmarksopen,bookmarksopenlevel=2,colorlinks,linkcolor=blue,citecolor=green]{hyperref}

\input{tcilatex}
\begin{document}

\title{\textbf{Hamiltonian Formalism of Two-Dimensional Vlasov Kinetic
Equation}}
\author{Maxim V.~Pavlov$^{1,2}$ \\
$^{1}$Sector of Mathematical Physics,\\
Lebedev Physical Institute of Russian Academy of Sciences,\\
Leninskij Prospekt 53, 119991 Moscow, Russia\\
$^{2}$Laboratory of Geometric Methods in Mathematical Physics,\\
Lomonosov Moscow State University,\\
Leninskie Gory 1, 119991 Moscow, Russia }
\date{}
\maketitle

\begin{abstract}
In this paper the two-dimensional Benney system describing long wave
propagation of a finite depth fluid motion and the multi-dimensional
Russo--Smereka kinetic equation describing a bubbly flow are considered. The
Hamiltonian approach established by J. Gibbons for one-dimensional Vlasov
kinetic equation is extended to a multi-dimensional case. A local
Hamiltonian structure associated with the hydrodynamic lattice of moments
derived by D.J. Benney is constructed. A relationship between this
hydrodynamic lattice of moments and the two-dimensional Vlasov kinetic
equation is found. In the two-dimensional case a Hamiltonian hydrodynamic
lattice for the Russo--Smereka kinetic model is constructed. Simple
hydrodynamic reductions are presented.
\end{abstract}

\tableofcontents

\textit{keywords}: collisionless kinetic equation, Hamiltonian structure,
hydrodynamic lattice, hydrodynamic reduction\textit{.}\noindent

\newpage

\ \ \ \ \ \ \ \ \ \ \ \ \ \ \ \ \ \ \ \ \ \ \ \ \ \ \ \ \ \ \ \ \ \ \ \ \ \
\ \ \ \ \ \ \ \ \ \ \ \ \ \ \ \ \ \ \ \ \ \ \ \ \ \ \ \ \ \ \ \ \ \ \ \ \ \ 
\textit{to honor of David J. Benney}

\section{Introduction}

Recently (see \cite{Benney2}) D.J. Benney considered a three-dimensional
motion of a finite depth fluid. Corresponding nonlinear system in partial
derivatives is written on three components of the velocity\footnote{%
according to the notation introduced by D.J. Benney, $x$ and $z$ are
horizontal Cartesian coordinates, $y$ is a vertical Cartesian coordinate; $u$
and $w$ are horizontal components of the velocity, $v$ is a vertical
component of the velocity.} $u(x,y,z,t),v(x,y,z,t),w(x,y,z,t)$ and the
profile $\eta (x,z,t)$ of a free surface\footnote{%
any lower index means a corresponding derivative.}:%
\begin{equation}
u_{x}+v_{y}+w_{z}=0,  \label{ben}
\end{equation}%
\begin{equation*}
u_{t}+uu_{x}+vu_{y}+wu_{z}=-\eta _{x},
\end{equation*}%
\begin{equation*}
w_{t}+uw_{x}+vw_{y}+ww_{z}=-\eta _{z}
\end{equation*}%
with the boundary conditions%
\begin{equation*}
v=0,\text{ \ }y=0,
\end{equation*}%
\begin{equation*}
\eta _{t}+u\eta _{x}+w\eta _{z}-v=0,\text{ \ }y=\eta .
\end{equation*}

Introducing infinitely many moments%
\begin{equation}
A^{k,m}(x,z,t)=\overset{\eta }{\underset{0}{\int }}%
u^{k}(x,y,z,t)w^{m}(x,y,z,t)dy,  \label{mom}
\end{equation}%
D.J. Benney derived a two-dimensional hydrodynamic chain or we shall call a
\textquotedblleft \textit{hydrodynamic lattice}\textquotedblright 
\begin{equation}
A_{t}^{k,m}+A_{x}^{k+1,m}+A_{z}^{k,m+1}+kA^{k-1,m}A_{x}^{0,0}+mA^{k,m-1}A_{z}^{0,0}=0,%
\text{ \ }k,m=0,1,...,  \label{3}
\end{equation}%
which has just four local conservation laws, i.e.%
\begin{equation*}
A_{t}^{0,0}+A_{x}^{1,0}+A_{z}^{0,1}=0,
\end{equation*}%
\begin{equation*}
A_{t}^{1,0}+\left( A^{2,0}+\frac{1}{2}(A^{0,0})^{2}\right)
_{x}+A_{z}^{1,1}=0,
\end{equation*}%
\begin{equation*}
A_{t}^{0,1}+A_{x}^{1,1}+\left( A^{0,2}+\frac{1}{2}(A^{0,0})^{2}\right)
_{z}=0,
\end{equation*}%
\begin{equation*}
(A^{0,2}+A^{2,0}+(A^{0,0})^{2})_{t}+(A^{3,0}+A^{1,2}+2A^{1,0}A^{0,0})_{x}+(A^{2,1}+A^{0,3}+2A^{0,1}A^{0,0})_{z}=0.
\end{equation*}

In next Section of this paper we present a local Hamiltonian structure%
\footnote{%
we assume summations with respect to each repeated index here and everywhere
below; any pure differential operator $\partial _{x},\partial _{z},\partial
_{p},\partial _{q}$ acts on all that is written in r.h.s.} of (\ref{3})%
\begin{equation}
A_{t}^{i,j}+[iA^{i+k-1,j+m}\partial _{x}+k\partial
_{x}A^{i+k-1,j+m}+jA^{i+k,j+m-1}\partial _{z}+m\partial _{z}A^{i+k,j+m-1}]%
\frac{\delta \mathbf{H}}{\delta A^{k,m}}=0,  \label{can}
\end{equation}%
consider simplest hydrodynamic reductions of this Benney hydrodynamic
lattice and construct a relationship with the two-dimensional Vlasov
(collisionless Boltzmann) kinetic equation (see, for instance, \cite{zakh})%
\begin{equation}
f_{t}+pf_{x}+qf_{z}-f_{p}A_{x}^{0,0}-f_{q}A_{z}^{0,0}=0.  \label{vlas}
\end{equation}%
Third Section of this paper is devoted to another two-dimensional Vlasov
kinetic equation appearing in description of a bubbly flow (see \cite{rs}).
Corresponding Hamiltonian hydrodynamic lattice also is found. Simplest
three-dimensional hydrodynamic reductions for both hydrodynamic lattices are
extracted.

\section{The Benney System}

In this paper we utilize the approach established in \cite{Gibbons} and
later significantly developed in \cite{GibRai} for a one-dimensional case. 
\textit{We show that this approach is useful also for a \textbf{%
multi-dimensional} case}. Without loss of generality and for simplicity we
restrict our consideration to two-dimensional case only.

\textbf{1}. Introducing moments\footnote{%
these integrals converge, for example, if $\lambda $ is bounded and 
\TEXTsymbol{\vert}$\lambda $\TEXTsymbol{\vert}$\rightarrow 0$ faster than
any $(\Sigma $\TEXTsymbol{\vert}$p_{s}$\TEXTsymbol{\vert})$^{-K}$, $\forall
K\geqslant 1$.}%
\begin{equation}
A^{k,m}(x,z,t)=\int \int p^{k}q^{m}f(x,z,p,q,t)dpdq,\text{ \ }k,m=0,1,...
\label{7}
\end{equation}%
Vlasov kinetic equation (\ref{vlas}) implies Benney hydrodynamic lattice%
\textit{\ }(\ref{3}).

\textbf{2}. Vlasov kinetic equation (\ref{vlas}) possesses the Hamiltonian
structure%
\begin{equation}
f_{t}=\{f,H\}_{\text{LP}},  \label{8}
\end{equation}%
where the standard (ultralocal) Lie--Poisson bracket%
\begin{equation*}
\{f,g\}_{\text{LP}}:=f_{p}g_{x}+f_{q}g_{z}-f_{x}g_{p}-f_{z}g_{q},
\end{equation*}%
and the Hamiltonian function $H=\frac{1}{2}(p^{2}+q^{2})+A^{0,0}$.

\textbf{3}. Introducing the functional%
\begin{equation}
\mathbf{H}=\frac{1}{2}\int \int (A^{0,2}+A^{0,2}+(A^{0,0})^{2})dxdz
\label{ham}
\end{equation}%
such that $H=\delta \mathbf{H}/\delta f$, Vlasov kinetic equation (\ref{vlas}%
) also can be written in the theoretic-field Hamiltonian form (cf. (\ref{8}))%
\begin{equation}
f_{t}=\{f,\mathbf{H}\},  \label{9}
\end{equation}%
where%
\begin{equation*}
\{f,\mathbf{H}\}:=(f_{p}\partial _{x}+f_{q}\partial _{z}-f_{x}\partial
_{p}-f_{z}\partial _{q})\frac{\delta \mathbf{H}}{\delta f}\mathbf{.}
\end{equation*}%
Indeed,%
\begin{equation*}
\frac{\delta \mathbf{H}}{\delta f}=\frac{1}{2}\frac{\delta A^{0,2}}{\delta f}%
+\frac{1}{2}\frac{\delta A^{2,0}}{\delta f}+A^{0,0}\frac{\delta A^{0,0}}{%
\delta f}=\frac{1}{2}(p^{2}+q^{2})+A^{0,0}\equiv H,
\end{equation*}%
where we utilized the elementary variational property (see (\ref{7}))%
\begin{equation*}
\frac{\delta A^{k,m}}{\delta f}=p^{k}q^{m}.
\end{equation*}%
Thus Vlasov kinetic equation (\ref{vlas}) is an integro-differential
equation (see (\ref{7})):%
\begin{equation}
f_{t}+pf_{x}+qf_{z}-f_{p}A_{x}^{0,0}-f_{q}A_{z}^{0,0}=0,\text{ \ }%
A^{0,0}(x,z,t)=\int \int f(x,z,p,q,t)dpdq.  \label{k}
\end{equation}

\textbf{4}. The relation between Vlasov kinetic equation (\ref{vlas})
written in Hamiltonian form (\ref{9}) and Benney hydrodynamic lattice (\ref%
{3}) written in Hamiltonian form (\ref{can}) is obtained by defining mapping%
\begin{equation*}
\mu :f(x,z,p,q,t)\longmapsto \{A^{k,m}(x,z,t)\}_{k,m=0,1,...}^{\infty },
\end{equation*}%
where moments $A^{k,m}$ are determined by (\ref{7}). Indeed, in a general
case (cf. (\ref{ham}))%
\begin{equation}
\mathbf{H}=\int \int
h(A^{0,0},A^{0,1},A^{1,0},A^{0,2},A^{1,1},A^{2,0},...)dxdz  \label{gen}
\end{equation}%
and%
\begin{equation*}
\frac{\delta \mathbf{H}}{\delta f}=\frac{\partial h}{\partial A^{k,m}}\frac{%
\delta A^{k,m}}{\delta f}=p^{k}q^{m}\frac{\partial h}{\partial A^{k,m}}.
\end{equation*}%
Then multiplying (\ref{9}) by $p^{k}q^{m}$ and integrating over $p$ and $q$
(see (\ref{7})),%
\begin{equation*}
A_{t}^{i,j}=\int \int p^{i}q^{j}f_{t}dpdq=\int \int p^{i}q^{j}\{f,\mathbf{H}%
\}dpdq
\end{equation*}%
\begin{equation*}
=\int \int p^{i}q^{j}\left[ f_{p}\left( \frac{\delta \mathbf{H}}{\delta f}%
\right) _{x}+f_{q}\left( \frac{\delta \mathbf{H}}{\delta f}\right)
_{z}-f_{x}\left( \frac{\delta \mathbf{H}}{\delta f}\right) _{p}-f_{z}\left( 
\frac{\delta \mathbf{H}}{\delta f}\right) _{q}\right] dpdq
\end{equation*}%
\begin{equation*}
=\left( \frac{\partial h}{\partial A^{k,m}}\right) _{x}\int \int
p^{k+i}q^{m+j}f_{p}dpdq+\left( \frac{\partial h}{\partial A^{k,m}}\right)
_{z}\int \int p^{k+i}q^{m+j}f_{q}dpdq
\end{equation*}%
\begin{equation*}
-k\frac{\partial h}{\partial A^{k,m}}\int \int p^{k+i-1}q^{m+j}f_{x}dpdq-m%
\frac{\partial h}{\partial A^{k,m}}\int \int p^{k+i}q^{m+j-1}f_{z}dpdq,
\end{equation*}%
then integrating by parts two first integrals (with respect to $p$ and $q$,
respectively), we obtain%
\begin{equation*}
=-(k+i)A^{k+i-1,m+j}\left( \frac{\partial h}{\partial A^{k,m}}\right)
_{x}-(m+j)A^{k+i,m+j-1}\left( \frac{\partial h}{\partial A^{k,m}}\right) _{z}
\end{equation*}%
\begin{equation*}
-k\frac{\partial h}{\partial A^{k,m}}A_{x}^{k+i-1,m+j}-m\frac{\partial h}{%
\partial A^{k,m}}A_{z}^{k+i,m+j-1}.
\end{equation*}%
This is nothing but precisely (\ref{can}), where obviously (see (\ref{gen}))%
\begin{equation*}
\frac{\delta \mathbf{H}}{\delta A^{k,m}}\equiv \frac{\partial h}{\partial
A^{k,m}}.
\end{equation*}

In particular case (\ref{ham}), Hamiltonian system (\ref{can}) yields Benney
hydrodynamic lattice (\ref{3}). Corresponding two-dimensional Poisson bracket%
\begin{equation*}
\{A^{i,j}(x,z),A^{k,m}(x^{\prime },z^{\prime
})\}=[(i+k)A^{i+k-1,j+m}\partial _{x}+kA_{x}^{i+k-1,j+m}
\end{equation*}%
\begin{equation}
+(j+m)A^{i+k,j+m-1}\partial _{z}+mA_{z}^{i+k,j+m-1}]\delta (x-x^{\prime
})\delta (z-z^{\prime })  \label{10}
\end{equation}%
we call the \textit{two-dimensional} Kupershmidt--Manin Poisson bracket,
because earlier (see \cite{KM}) the \textit{one-dimensional}
Kupershmidt--Manin Poisson bracket%
\begin{equation}
\{A^{k}(x),A^{m}(x^{\prime })\}=[(k+m)A^{k+m-1}\partial
_{x}+mA_{x}^{k+m-1}]\delta (x-x^{\prime })  \label{km}
\end{equation}%
was derived for the one-dimensional Benney hydrodynamic chain (see \cite%
{Benney})%
\begin{equation*}
A_{t}^{k}+A_{x}^{k+1}+kA^{k-1}A_{x}^{0}=0,\text{ }k=0,1,...
\end{equation*}%
Poisson bracket (\ref{10}) is a \textit{first} example of \textit{%
two-dimensional} differential-geometric Poisson brackets of a first order
(see \cite{dn}, \cite{mokh} and \cite{fls}) generalised to an \textit{%
infinitely} many component case.

In the previous Section we mentioned that D.J. Benney was able to find just
four local conservation laws. Actually, explanation of this result is based
on existence of local Hamiltonian structure (\ref{can}). Indeed, any
Hamiltonian system (\ref{can}) possesses just four local conservation laws
for an arbitrary Hamiltonian density (\ref{gen}):

\textbf{1}. the continuity conservation law%
\begin{equation*}
A_{t}^{0,0}+\left( kA^{k-1,m}\frac{\partial h}{\partial A^{k,m}}\right)
_{x}+\left( mA^{k,m-1}\frac{\partial h}{\partial A^{k,m}}\right) _{z}=0;
\end{equation*}

\textbf{2}. the conservation law of the momentum ($x,z$ components)%
\begin{equation*}
A_{t}^{1,0}+\left[ (k+1)A^{k,m}\frac{\partial h}{\partial A^{k,m}}-h\right]
_{x}+\left( mA^{k+1,m-1}\frac{\partial h}{\partial A^{k,m}}\right) _{z}=0,
\end{equation*}%
\begin{equation*}
A_{t}^{0,1}+\left( kA^{k-1,m+1}\frac{\partial h}{\partial A^{k,m}}\right)
_{x}+\left[ (m+1)A^{k,m}\frac{\partial h}{\partial A^{k,m}}-h\right] _{z}=0;
\end{equation*}

\textbf{3}. the conservation law of the energy%
\begin{equation*}
h_{t}+\left( kA^{i+k-1,j+m}\frac{\partial h}{\partial A^{i,j}}\frac{\partial
h}{\partial A^{k,m}}\right) _{x}+\left( mA^{i+k,j+m-1}\frac{\partial h}{%
\partial A^{i,j}}\frac{\partial h}{\partial A^{k,m}}\right) _{z}=0.
\end{equation*}

Any other local conservation laws can exist just for very special
dependencies of Hamiltonian density $%
h(A^{0,0},A^{0,1},A^{1,0},A^{0,2},A^{1,1},A^{2,0},...)$. We believe that
corresponding hydrodynamic lattices must be integrable. This problem should
be considered in a separate paper.

The construction presented in this Section is so natural, that without any
restrictions, one can generalise two-dimensional Kupershmidt--Manin Poisson
bracket (\ref{10}) to any higher number of spatial dimensions.

\section{The Russo--Smereka Model}

The Russo--Smereka model (see \cite{rs}) describing a bubbly flow contains
the Vlasov kinetic equation%
\begin{equation}
f_{t}+u^{k}\frac{\partial f}{\partial x^{k}}-\frac{\partial (p_{m}u^{m})}{%
\partial x^{k}}\frac{\partial f}{\partial p_{k}}=0,  \label{kin}
\end{equation}%
where ($\rho _{l}$ and $\tau $ are parameters, indices $k,m$ run from $1$ up
to $n$)%
\begin{equation*}
\rho _{l}u^{k}=\frac{2}{\tau }p_{k}+3(j_{k}-3\partial _{k}\Phi ),\text{ \ }%
j_{k}=\int p_{k}fd\mathbf{p}
\end{equation*}%
and the Poisson equation%
\begin{equation}
\partial _{m}^{2}\Phi =\partial _{m}j_{m}.  \label{poi}
\end{equation}%
In this Section we adopt the approach established in \cite{Gibbons} and
developed in \cite{GibRai} to the case equipped by extra constraints.

\textbf{Theorem}: \textit{Vlasov kinetic equation} (\ref{kin}) \textit{and
Poisson equation} (\ref{poi}) \textit{can be written in the theoretic-field
Hamiltonian form (cf.} (\ref{9}))%
\begin{equation}
f_{t}=\{f,\mathbf{H}\},\text{ \ }\frac{\delta \mathbf{H}}{\delta \Phi }=0,
\label{hami}
\end{equation}%
\textit{where the Poisson bracket is}%
\begin{equation*}
\{f,\mathbf{H}\}:=\left( \frac{\partial f}{\partial p_{m}}\frac{\partial }{%
\partial x^{m}}-\frac{\partial f}{\partial x^{m}}\frac{\partial }{\partial
p_{m}}\right) \frac{\delta \mathbf{H}}{\delta f}
\end{equation*}%
\textit{and the Hamiltonian is}%
\begin{equation}
\mathbf{H}=\frac{9}{2\rho _{l}}\overset{n}{\underset{k=1}{\sum }}\int \left( 
\frac{2}{9\tau }\int p_{k}^{2}fd\mathbf{p}+\frac{1}{3}j_{k}^{2}-2j_{k}%
\partial _{k}\Phi +(\partial _{k}\Phi )^{2}\right) d\mathbf{x.}  \label{gami}
\end{equation}

\textbf{Proof}: Substitution Hamiltonian (\ref{gami}) into (\ref{hami})
yields (\ref{kin}) and (\ref{poi}), respectively.

Without loss of generality and for simplicity here we again restrict our
consideration on two-dimensional case only. Then Hamiltonian system (\ref%
{hami})%
\begin{equation*}
f_{t}=(f_{p}\partial _{x}+f_{q}\partial _{z}-f_{x}\partial
_{p}-f_{z}\partial _{q})\frac{\delta \mathbf{H}}{\delta f},\text{ \ }\frac{%
\delta \mathbf{H}}{\delta \Phi }=0
\end{equation*}%
is determined by the Hamiltonian (\ref{gami})%
\begin{equation*}
\mathbf{H}=\frac{9}{2\rho _{l}}\int \left( \frac{2}{9\tau }(A^{0,2}+A^{2,0})+%
\frac{1}{3}[(A^{0,1})^{2}+(A^{1,0})^{2}]-2(A^{1,0}\Phi _{x}+A^{0,1}\Phi
_{z})+\Phi _{x}^{2}+\Phi _{z}^{2}\right) dxdz.
\end{equation*}%
Corresponding Hamiltonian hydrodynamic lattice (\ref{can}) takes the form
(cf. (\ref{3}); here both parameters $\rho _{l}$ and $\tau $ removed by
appropriate scaling of independent variables $x,z,t$ and dependent functions 
$A^{k,m}$ and $\Phi $)%
\begin{equation}
0=A_{t}^{i,j}+A_{z}^{i,j+1}+A_{x}^{i+1,j}  \label{rs}
\end{equation}%
\begin{equation*}
+[iA^{i-1,j+1}\partial _{x}+(j+1)A^{i,j}\partial
_{z}+A_{z}^{i,j}](A^{0,1}-3\Phi _{z})+[(i+1)A^{i,j}\partial
_{x}+A_{x}^{i,j}+jA^{i+1,j-1}\partial _{z}](A^{1,0}-3\Phi _{x}),
\end{equation*}%
where Poisson equation (\ref{poi}) becomes%
\begin{equation*}
\Phi _{xx}+\Phi _{zz}=A_{x}^{1,0}+A_{z}^{0,1}.
\end{equation*}%
This constraint $\delta \mathbf{H}/\delta \Phi =0$ can be written in the
obvious conservative form $\partial _{m}(\partial _{m}\Phi -j_{m})=0$. Thus
in the two-dimensional case one extra conservation law exists, i.e.%
\begin{equation}
(\Phi _{x}-A^{1,0})_{x}+(\Phi _{z}-A^{0,1})_{z}=0,  \label{son}
\end{equation}%
while the hydrodynamic lattice possesses just four local conservation laws
(cf. two previous Sections):

\textbf{1}. the continuity conservation law%
\begin{equation*}
A_{t}^{0,0}+(A^{1,0}+A^{0,0}A^{1,0}-3A^{0,0}\Phi
_{x})_{x}+(A^{0,1}+A^{0,0}A^{0,1}-3A^{0,0}\Phi _{z})_{z}=0,
\end{equation*}

\textbf{2}. the conservation law of the momentum ($x,z$ components)%
\begin{equation*}
A_{t}^{0,1}+(A^{1,1}+A^{0,1}A^{1,0}-3A^{0,1}\Phi _{x}-3A^{1,0}\Phi
_{z}+3\Phi _{x}\Phi _{z})_{x}
\end{equation*}%
\begin{equation*}
+\left( A^{0,2}+\frac{3}{2}(A^{0,1})^{2}+\frac{1}{2}(A^{1,0})^{2}-6A^{0,1}%
\Phi _{z}+\frac{3}{2}\Phi _{z}^{2}-\frac{3}{2}\Phi _{x}^{2}\right) _{z}=0,
\end{equation*}%
\begin{equation*}
A_{t}^{1,0}+(A^{1,1}+A^{0,1}A^{1,0}-3A^{0,1}\Phi _{x}-3A^{1,0}\Phi
_{z}+3\Phi _{x}\Phi _{z})_{z}
\end{equation*}%
\begin{equation*}
+\left( A^{2,0}+\frac{1}{2}(A^{0,1})^{2}+\frac{3}{2}(A^{1,0})^{2}-6A^{1,0}%
\Phi _{x}+\frac{3}{2}\Phi _{x}^{2}-\frac{3}{2}\Phi _{z}^{2}\right) _{x}=0,
\end{equation*}

\textbf{3}. the conservation law of the energy%
\begin{equation*}
0=(A^{0,2}+A^{2,0}+(A^{0,1})^{2}+(A^{1,0})^{2}-6A^{1,0}\Phi
_{x}-6A^{0,1}\Phi _{z}+3\Phi _{x}^{2}+3\Phi _{z}^{2})_{t}
\end{equation*}%
\begin{equation*}
+(A^{3,0}+A^{1,2}+A^{1,0}A^{0,2}+3A^{2,0}A^{1,0}-9A^{2,0}\Phi
_{x}-3A^{0,2}\Phi _{x}+2A^{0,1}A^{1,1}+2(A^{1,0})^{3}+2(A^{0,1})^{2}A^{1,0}
\end{equation*}%
\begin{equation*}
+18A^{1,0}\Phi _{x}^{2}+18A^{0,1}\Phi _{x}\Phi _{z}+6(A^{1,0}-\Phi _{x})\Phi
_{t}-6(A^{1,1}+A^{0,1}A^{1,0})\Phi _{z}-6[(A^{0,1})^{2}+2(A^{1,0})^{2}]\Phi
_{x})_{x}
\end{equation*}%
\begin{equation*}
+(A^{2,1}+A^{0,3}+A^{0,1}A^{2,0}+3A^{0,2}A^{0,1}-9A^{0,2}\Phi
_{z}-3A^{2,0}\Phi _{z}+2A^{1,0}A^{1,1}+2(A^{0,1})^{3}+2(A^{1,0})^{2}A^{0,1}
\end{equation*}%
\begin{equation*}
+18A^{0,1}\Phi _{z}^{2}+18A^{1,0}\Phi _{x}\Phi _{z}+6(A^{0,1}-\Phi _{z})\Phi
_{t}-6(A^{1,1}+A^{0,1}A^{1,0})\Phi _{x}-6[(A^{1,0})^{2}+2(A^{0,1})^{2}]\Phi
_{z})_{z}.
\end{equation*}

\section{Multi-Component Reductions}

Investigation of infinitely many component quasilinear systems of a first
order (i.e. hydrodynamic lattices) is very complicated problem. One of most
effective tools is a method of multi-component hydrodynamic reductions. In a
general case hydrodynamic reductions can be extracted utilizing \textit{%
generalized} functions (see, for instance, \cite{chesn}) like the Heaviside
step-function or the Dirac delta-function.

For example, the so called \textquotedblleft cold plasma\textquotedblright\
approximation ansatz\footnote{%
also well known as the \textquotedblleft multi-flow\textquotedblright\
ansatz.} (see, for instance, \cite{Silin})%
\begin{equation*}
f(x,z,p,q,t)=\underset{k=1}{\overset{N}{\sum }}\eta ^{k}(x,z,t)\delta
(p-u^{k}(x,z,t))\delta (q-w^{k}(x,z,t))
\end{equation*}%
reduces Benney hydrodynamic lattice (\ref{3}) to the finite component form%
\begin{equation*}
\eta _{t}^{k}+(u^{k}\eta ^{k})_{x}+(w^{k}\eta ^{k})_{z}=0,
\end{equation*}%
\begin{equation*}
u_{t}^{k}+u^{k}u_{x}^{k}+w^{k}u_{z}^{k}+\left( \underset{m=1}{\overset{N}{%
\sum }}\eta ^{m}\right) _{x}=0,\text{ \ }%
w_{t}^{k}+u^{k}w_{x}^{k}+w^{k}w_{z}^{k}+\left( \underset{m=1}{\overset{N}{%
\sum }}\eta ^{m}\right) _{z}=0,
\end{equation*}%
where all moments are expressed polynomially via new field variables $\eta
^{k},u^{s},w^{m}$ (see (\ref{7}) and more detail in \cite{MaksHam})%
\begin{equation}
A^{k,m}=\underset{p=1}{\overset{N}{\sum }}\eta ^{p}(u^{p})^{k}(w^{p})^{m}.
\label{zakh}
\end{equation}

In the simplest case $N=1$, this is nothing but a system describing the
irrotational two-dimensional hydrodynamics%
\begin{equation}
\eta _{t}+(u\eta )_{x}+(w\eta )_{z}=0,  \label{irro}
\end{equation}%
\begin{equation*}
u_{t}+uu_{x}+wu_{z}+\eta _{x}=0,\text{ \ }w_{t}+uw_{x}+ww_{z}+\eta _{z}=0.
\end{equation*}

\textbf{1}. This system also can be obtained directly from two-dimensional
Benney system (\ref{ben}) by the reduction\footnote{%
substitution reduced dependencies $u(x,z,t)$ and $w(x,z,t)$ into (\ref{mom})
transforms hydrodynamic lattice (\ref{3}) to above three component two
dimensional hydrodynamic type system (\ref{irro}).} $u(x,z,t),w(x,z,t)$ and $%
v=-y(u_{x}+w_{z})$.

\textbf{2}. This system possesses a local Hamiltonian structure (see, for
instance, \cite{MaksMZ}), where corresponding Poisson bracket is (here we
write just nonzero components)%
\begin{equation*}
\{\eta (x,z),u(x^{\prime },z^{\prime })\}=\{u(x,z),\eta (x^{\prime
},z^{\prime })\}=\delta ^{\prime }(x-x^{\prime })\delta (z-z^{\prime }),
\end{equation*}%
\begin{equation*}
\{\eta (x,z),w(x,z)\}=\{w(x,z),\eta (x^{\prime },z^{\prime })\}=\delta
(x-x^{\prime })\delta ^{\prime }(z-z^{\prime }),
\end{equation*}%
\begin{equation*}
\{u(x,z),w(x^{\prime },z^{\prime })\}=-\{w(x,z),u(x^{\prime },z^{\prime })\}=%
\frac{u_{z}-w_{x}}{\eta }\delta (x-x^{\prime })\delta (z-z^{\prime }).
\end{equation*}%
Indeed, the substitution $A^{k,m}=\eta u^{k}w^{m}$ into two-dimensional
Kupershmidt--Manin Poisson bracket (\ref{10}) implies above two-dimensional
Poisson bracket.

\textbf{3}. This system possesses just four local conservation laws (see the
end of the previous Section)%
\begin{equation*}
\eta _{t}+(u\eta )_{x}+(w\eta )_{z}=0,
\end{equation*}%
\begin{equation*}
(u\eta )_{t}+\left( u^{2}\eta +\frac{1}{2}\eta ^{2}\right) _{x}+(uw\eta
)_{z}=0,
\end{equation*}%
\begin{equation*}
(w\eta )_{t}+(uw\eta )_{x}+\left( w^{2}\eta +\frac{1}{2}\eta ^{2}\right)
_{z}=0,
\end{equation*}%
\begin{equation*}
\lbrack (u^{2}+w^{2})\eta +\eta ^{2}]_{t}+[(u^{2}+w^{2})u\eta +2u\eta
^{2}]_{x}+[(u^{2}+w^{2})w\eta +2w\eta ^{2}]_{z}=0.
\end{equation*}

In comparison with hydrodynamic lattice (\ref{3}) the substitution (\ref%
{zakh}) into hydrodynamic lattice (\ref{rs}) leads to another
multi-component three-dimensional quasilinear system of a first order%
\footnote{%
also first two equations of this quasilinear system are nothing but the
Poisson equation written in conservative form (\ref{son}).} (here we
introduced two extra field variables $a=\Phi _{x}$ and $b=\Phi _{z}$)%
\begin{equation*}
\eta _{t}^{k}+[\eta ^{k}(A^{1,0}-3a+u^{k})]_{x}+[\eta
^{k}(A^{0,1}-3b+w^{k})]_{z}=0,
\end{equation*}%
\begin{equation*}
u_{t}^{k}+u^{k}u_{x}^{k}+w^{k}u_{z}^{k}+[w^{k}\partial
_{x}+u_{z}^{k}](A^{0,1}-3b)+[u^{k}(A^{1,0}-3a)]_{x}=0,
\end{equation*}%
\begin{equation*}
w_{t}^{k}+w^{k}w_{z}^{k}+u^{k}w_{x}^{k}+[w^{k}(A^{0,1}-3b)]_{z}+[w_{x}^{k}+u^{k}\partial _{z}](A^{1,0}-3a)=0,
\end{equation*}%
\begin{equation*}
a_{z}=b_{x},\text{ \ }(a-A^{1,0})_{x}+(b-A^{0,1})_{z}=0,
\end{equation*}%
where two last equations follow from the constraint $\delta \mathbf{H}%
/\delta \Phi =0$ (see (\ref{son})).

Obviously any solutions of above finite-component quasilinear systems of a
first order are simultaneously solutions of hydrodynamic lattices (\ref{3}),
(\ref{rs}). Construction of these solutions should be made in a separate
paper.

\section{Conclusion}

In this paper we extended the Hamiltonian approach established in \cite%
{Gibbons} and developed in \cite{GibRai} to a multi-dimensional case and to
the case equipped by extra constraints, which can be written in a
variational form. We constructed Hamiltonian structure (\ref{can}) for
Benney hydrodynamic lattice (\ref{3}) associated with two-dimensional Benney
system (\ref{ben}) and we established a link between two-dimensional Benney
system (\ref{ben}) and two-dimensional Vlasov kinetic equation (\ref{vlas}),
which is an integro-differential equation (\ref{k}). We constructed
hydrodynamic lattice (\ref{rs}) associated with two-dimensional
Russo--Smereka kinetic equation (see (\ref{kin})), and we found a
Hamiltonian structure of this hydrodynamic lattice. Also, simple
three-dimensional hydrodynamic reductions are presented.

Here we just mention how the construction presented in this paper can be
utilized. If, for instance, instead of the Hamiltonian density (see (\ref%
{ham}))%
\begin{equation*}
h=\frac{1}{2}[A^{0,2}+A^{2,0}+(A^{0,0})^{2}]
\end{equation*}%
to substitute a slightly more general expression (here $Q(a)$ is an
arbitrary function)%
\begin{equation*}
h=\frac{1}{2}[A^{0,2}+A^{2,0}+Q(A^{0,0})]
\end{equation*}%
into (\ref{can}), one can obtain the hydrodynamic lattice%
\begin{equation*}
A_{t}^{k,m}+A_{x}^{k+1,m}+A_{z}^{k,m+1}+kA^{k-1,m}Q^{\prime \prime
}(A^{0,0})A_{x}^{0,0}+mA^{k,m-1}Q^{\prime \prime }(A^{0,0})A_{z}^{0,0}=0,%
\text{ \ }k,m=0,1,...,
\end{equation*}%
whose three component hydrodynamic reduction determined by the moment
decomposition $A^{k,m}=\eta u^{k}w^{m}$ is the nonlinear system describing
irrotational two-dimensional barotropic hydrodynamics (cf. (\ref{irro}))%
\begin{equation*}
\eta _{t}+(u\eta )_{x}+(w\eta )_{z}=0,
\end{equation*}%
\begin{equation*}
u_{t}+uu_{x}+wu_{z}+Q^{\prime \prime }(\eta )\eta _{x}=0,\text{ \ }%
w_{t}+uw_{x}+ww_{z}+Q^{\prime \prime }(\eta )\eta _{z}=0.
\end{equation*}%
Thus this nonlinear system (as in the previous Section) is the reduction $%
u(x,z,t),w(x,z,t)$ and $v=-y(u_{x}+w_{z})$ of the barotropic
\textquotedblleft Benney type\textquotedblright\ fluid (cf. (\ref{ben}))%
\begin{equation*}
u_{x}+v_{y}+w_{z}=0,
\end{equation*}%
\qquad \qquad \qquad 
\begin{equation*}
u_{t}+uu_{x}+vu_{y}+wu_{z}+Q^{\prime \prime }(\eta )\eta _{x}=0,
\end{equation*}%
\begin{equation*}
w_{t}+uw_{x}+vw_{y}+ww_{z}+Q^{\prime \prime }(\eta )\eta _{z}=0
\end{equation*}%
with the boundary conditions%
\begin{equation*}
v=0,\text{ \ }y=0,
\end{equation*}%
\begin{equation*}
\eta _{t}+u\eta _{x}+w\eta _{z}-v=0,\text{ \ }y=\eta .
\end{equation*}

More complicated Hamiltonian hydrodynamic lattices and more complicated
hydrodynamic reductions (see, for instance, \cite{chesn}) should be
investigated in a separate paper.

We hope that two-dimensional Kupershmidt--Manin Poisson bracket (\ref{10})
will play the same important role in the theory of integrable hydrodynamic
lattices as well as one-dimensional Kupershmidt--Manin Poisson bracket (\ref%
{km}) in the theory of integrable hydrodynamic chains (see, for instance, 
\cite{MaksHam} and \cite{FerMarsh}).

\section*{Acknowledgements}

Author thanks Alexander Chesnokov, Sergey Gavrilyuk, Oleg Mokhov, Vladimir
Taranov, Sergey Tsarev and Victor Vedenyapin for their stimulating and
clarifying discussions.

MVP's work was partially supported by the RF Government grant
\#2010-220-01-077, ag. \#11.G34.31.0005, by the grant of Presidium of RAS
\textquotedblleft Fundamental Problems of Nonlinear
Dynamics\textquotedblright\ and by the RFBR grant 14-01-00389.

\end{document}